\documentstyle[aps,psfig]{revtex}
\newcommand{\be}{\begin{equation}} 
\newcommand{\ee}{\end{equation}} 
\newcommand{\ben}{\begin{eqnarray}} 
\newcommand{\een}{\end{eqnarray}} 
\begin{document} 

\title{Soliton model for proton conductivity in Langmuir films} 
\author{D. Bazeia{$^1$}, V. B. P. Leite{$^{2}$}\footnote{Corresponding
author. Fax: +55 17 221 2247. E-mail: {\it vleite@df.ibilce.unesp.br}},
B. H. B. Lima{$^3$}, and F. Moraes{$^3$}}
\address{{$^1$}Departamento de F\'\i sica, Universidade Federal da
Para\'{\i}ba\\ Caixa Postal 5008, 58051-970 Jo\~ao Pessoa, PB, Brazil\\
{$^2$}Departamento de F\'\i sica, IBILCE, Universidade Estadual Paulista\\
15054-000 S\~ao Jos\'e do Rio Preto, SP, Brazil\\
{$^3$}Laborat\'orio de F\'\i sica  Te\'orica e Computacional, Departamento
de F\'\i sica, \\Universidade Federal de Pernambuco,
50670-901 Recife, PE, Brazil}
\date{\today} 
\maketitle 

\begin{abstract} 
A soliton model for proton conductivity in Langmuir films is
presented. The model contains three real scalar fields describing 
the hydrogen involved in the conduction, the hydrophilic
head of the Langmuir film, and the water. Soliton solutions that
describe proton motion along the hydrogen bonds are found. Under
compression of the film, the distance between the minima of the proton
potential and the strength of the hydrogen bonds between the film
molecule and the water are changed. Such changes increase the
probability of soliton creation.  The model presented allows proton
conductivity data in Langmuir films to be explained.\\
\end{abstract}  

\vspace{2cm}

Research on the characterization of Langmuir films has led to the
identification of a critical packing density, below which some properties
of the system change significantly \cite{e1,e1a,e2,e3,e4,e4a,e5}.
In particular, measurements of lateral conductance in Langmuir films 
suggested the authors of Ref.~{\cite{e6,e6a}}
to postulate the conductance arises from proton
transfer via the presence of hydrogen bonds in the film network. The
likelihood of this assumption is supported by the fact that in most
of such films there is no room for ionic or electronic conduction.

There are few theoretical efforts to understand
the mechanisms behind such protonic conductance; see Ref.~{\cite{lco98}}
and references therein. In Ref.~{\cite{lco98}} the issue is addressed via
a simple unidimensional mechanism for proton transfer in the hydrogen bonds.
In this model, the two oxygen atoms involved in the proton transfer are
treated equivalently. This is an approximation, because one oxygen belongs
to a water molecule, and the other to a hydrophilic headgroup.

In the present work another model is proposed based on the soliton models
of Refs.~{\cite{pne88,xzh96,bnt97}}. The model of Ref.~{\cite{pne88}} deals
with proton conduction in quasi-unidimensional networks that arise due to
hydrogen bonds. This model corresponds to a two-component chain, the proton
subchain and the heavy nuclei subchain connected by the hydrogen bonds. 
It describes two real scalar fields, $\phi=\phi(x,t)$ and $\chi=\chi(x,t)$,
in two-dimensional space-time. These fields correspond to the lighter
($\phi$) and heavier ($\chi$) subchains, respectively. The interactions
between the two sublattices is described by the potential $U=U(\phi,\chi)$,
a functional of the two fields.

A related problem to the two-component chain is the diatomic one, which is
modeled in Ref.~{\cite{xzh96}}. The model introduces a derivative
coupling between the two elements that form the diatomic chain.
This model was further investigated in Ref.~{\cite{bnt97}}, where ideas
of \cite{pne88,xzh96,first} are combined to incorporate the
derivative coupling between the lighter and heavier subchains of the
two-component model. It was used to investigate hydrogen-bonded systems.
The model describes quasi-unidimensional systems that can be represented
by repetitions of the basic entity $X-H\cdots$, where $H$ and $X$ represent
the proton and the heavy nuclei, and $-$ and $\cdots$ the covalent and the
hydrogen bond, respectively.  The simplest example is water \cite{pne88},
where $X$ stands for $OH$. We can think of one-component and two-component
models, but the two-component model is more realistic, because the network
is visualized as a diatomic chain, composed of two sublattices, the lighter
sublattice describing the protons, and the heavier sublattice the heavier
component of the system. Langmuir films are spread and compressed on an
aqueous subphase (substrate in our case), and we recall that the experimental
measurements in such films are in general done at room temperature. The film
is considered an isolated system, having no interference from the water
substrate. The water substrate only provides hydration water that participates
in the H-bond network. In the model, we do not have to assusme any specific
crystaline structure for the hydrated water.

The main goal of the present work is to offer a new model to understand
the proton conductivity problem in Langmuir films. This model appears through
the hydrogen bonds that connect the hydrophilic head of the film molecules
to the water, leading to a mechanism that allows the presence of solitons,
that drive the proton motion in the network. To do this, we first extend
the model of Ref.~{\cite{bnt97}} to the case of a three-component
chain, which seems to be more appropriate to the investigation of such films.
The three-component network in the Langmuir films can be schematically
represented by repetitions of $X-H\cdots Y-H\cdots$. It is composed of
protons and the $X$ and $Y$ groups that represent $OH$ and the amphiphilic
molecule that characterizes the film, respectively. They are spatially
represented in Fig.~1 in the case of the aliphatic acid. 

We justify a three-component model for such system recalling that
even when one neglects the tail contribution to the motion of the head,
it is much heavier than the water counterpart. The proton subchain is much
lighter than the other two subchains, so we consider the lighter sublattice
coupled with each one of the two other sublattices. We assume
the two heavier sublattices do not interact with one another by any
other mechanism. That is, the film molecules only interaction with the water
is via the hydrogen bonds.

Our model uses three real scalar
fields $\phi,\chi,$ and $\psi$, in bidimensional space-time. $\phi=\phi(x,t)$
describes translational motion of protons, and $\chi=\chi(x,t)$
and $\psi=\psi(x,t)$ the motion of $OH$ groups and amphiphilic molecule
subchains, respectively. We use standard notation, with $x^{\alpha}=(t,x)$,
$x_{\alpha}=(t,-x)$, and $\hbar=c=1$. The Lagrangian density is given by 
\ben
{\cal L}&=&\frac{1}{2}\,\partial_{\alpha}\phi\partial^{\alpha}\phi+
\frac{1}{2}\,{\mu}_{1}\,\partial_{\alpha}\chi\partial^{\alpha}\chi+
\frac{1}{2}\,{\mu}_{2}\,\partial_{\alpha}\psi\partial^{\alpha}\psi
\nonumber\\
& &+{{\nu}_{1}}\,\partial_{\alpha}\phi\partial^{\alpha}\chi+
{{\nu}_2}\,\partial_{\alpha}\phi\partial^{\alpha}\psi-U(\phi,\chi,\psi).
\een
The parameters $\mu_1$, $\mu_2$, and $\nu_1$, $\nu_2$ are real and positive,
and $U=U(\phi,\chi,\psi)$ is the potential. Here we consider the protons as
unit mass particles, so the parameters $\mu_1$ and $\mu_2$ can represent the
mass ratio between the $OH$ group and the hydrogen, and the amphiphilic
molecule and the hydrogen. The other two parameters $\nu_1$ and $\nu_2$
describe the derivative coupling between the proton and water and the
proton and amphiphilic molecule, respectively.  

The equations of motion are
\be
\label{gm1}
\frac{d^2\phi}{dx^2}+{\nu}_1\frac{d^2\chi}{dx^2}+
{\nu}_2\frac{d^2\psi}{dx^2}=\frac{\partial U}{\partial\phi},
\ee

\be
\label{gm2}
{\mu}_1\frac{d^2\chi}{dx^2}+{\nu}_1\frac{d^2\phi}{dx^2}=
\frac{\partial U}{\partial\chi},
\ee
and
\be
\label{gm3}
{\mu}_2\frac{d^2\psi}{dx^2}+{\nu}_2\frac{d^2\phi}{dx^2}=
\frac{\partial U}{\partial\psi}.
\ee
We consider the case in which the potential is defined
by some smooth function $W=W(\phi,\chi,\psi)$, in the form
\be
U(\phi,\chi,\psi)=\frac{1}{2}W^2_{\phi}+\frac{1}{2}W^2_{\chi}+
\frac{1}{2}W^2_{\psi}.
\ee
$W_{\phi}$ stands for $\partial W/\partial\phi,$ and so forth.
This specific form of the potential, together with ideas first introduced
in Refs.~{\cite{ope,ope1,ope2}} have been recently used to model topological
twistons in crystalline polyethylene \cite{pe,pe1}. Other investigations
related to the subject can be found in
Refs.~{\cite{susy,s1,bbr00,bbr001}} and in references therein.

%%%%%%%%%%%%%%%%%%%%%%%%%% FIGURE %%%%%%%%%%%%%%%%%%%%%%%%%%%%%%%%%% 
\begin{figure} 
\centerline{\psfig{figure=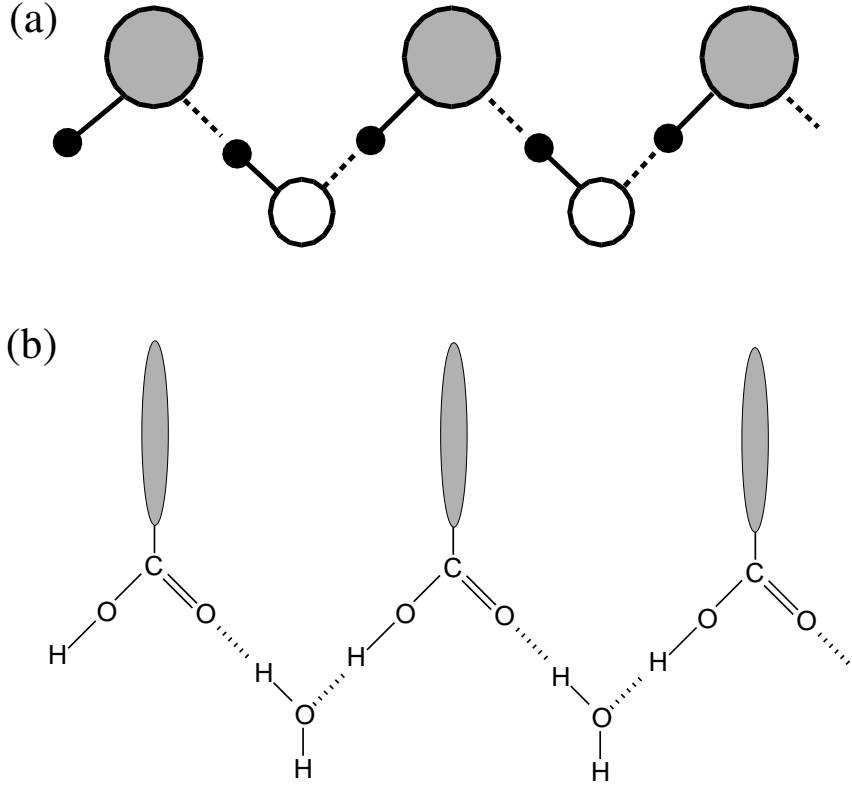,width=12.0cm}} 
\vspace{1.0cm} 
\caption{Schematic view of the three-component network. (a)  The proton,
the $OH$ group, and the head of the film, are represented by small, medium
and large circles, respectively. (b) Shows explicitly the atoms of the
diagram (a). The hydrophobic tail of the amphiphilic molecule is
represented by the gray area.} 
\end{figure}
%%%%%%%%%%%%%%%%%%%%%%%%%%%%%%%%%%%%%%%%%%%%%%%%%%%%%%%%%%%%%%%%%%%%%

In Ref.~{\cite{bnt97}} the derivative coupling between the two heavier
subchains was considered in the two-component model there studied. In the
present work we are considering the three-component model just introduced,
and no coupling between the two heavier subchains is taken into account,
since they seem to couple mainly through the hydrogen bonds.
We follow this reasoning and neglect further couplings between the $\chi$
and $\psi$ fields in the potential. Also the complexity of the problem
can be reduced by assuming the heavier subchains contribute insignificantly
to the potential, at least when compared to the lighter subchain. Therefore,
in this approximation $U=U(\phi)$ depends only on the mobility of the proton
field, which yields to
\be
U(\phi)=\frac{1}{2}W_{\phi}^2.
\ee
The equations of motion for static field configurations are reduced to
\be
\label{sm1}
\frac{d^2\phi}{dx^2}+{\nu}_1\frac{d^2\chi}{dx^2}+
{\nu}_2\frac{d^2\psi}{dx^2}=W_{\phi}W_{\phi\phi},
\ee

\be
\label{sm2}
{\mu}_1\frac{d^2\chi}{dx^2}+{\nu}_1\frac{d^2\phi}{dx^2}=0,
\ee
and
\be
\label{sm3}
{\mu}_2\frac{d^2\psi}{dx^2}+{\nu}_2\frac{d^2\phi}{dx^2}=0.
\ee
We consider fields whose derivatives present similar asymptotic behavior.
In this case the above equations (\ref{sm2}) and (\ref{sm3}) change to,
after setting to zero the integration constants,
\ben
\label{chi}
\frac{d\chi}{dx}=-\,\frac{\nu_1}{\mu_1}\,\frac{d\phi}{dx},
\\
\label{psi}
\frac{d\psi}{dx}=-\,\frac{\nu_2}{\mu_2}\,\frac{d\phi}{dx}.
\een

For static configurations the non-vanishing components of the energy-momentum
tensor $T^{\alpha\beta}$ can be written as $T^{00}=T_g+U(\phi)$
and $T^{11}=T_g-U(\phi)$, where $T_g$ represents the gradient contribution.
This is given by
\ben
T_g&=&\frac{1}{2}\Biggl[\,
\left(\frac{d\phi}{dx}\right)^2+
{\mu}_1\left(\frac{d\chi}{dx}\right)^2+
{\mu}_2\left(\frac{d\psi}{dx}\right)^2\nonumber\\
& &+2\,{\nu}_1\left(\frac{d\phi}{dx}\right)\left(\frac{d\chi}{dx}\right)+
2\,{\nu}_2\left(\frac{d\phi}{dx}\right)\left(\frac{d\psi}{dx}\right)\,\Biggr].
\een
We use the solutions (\ref{chi}) and (\ref{psi}) to write
\ben
T^{00}&=&\frac{1}{2}\,\left[B^2\,\left(\frac{d\phi}{dx}\right)^2+
\left(\frac{dW}{d\phi}\right)^2\right],
\\
T^{11}&=&\frac{1}{2}\,\left[B^2\,\left(\frac{d\phi}{dx}\right)^2-
\left(\frac{dW}{d\phi}\right)^2\right],
\een
 where $B^2=1-\nu^2_1/\mu_1-\nu^2_2/\mu_2$. $B$ is considered
real and positive.

The energy density of the static solutions $\varepsilon(x)$
is  identified with $T^{00}$, and it can be written as
\be
\varepsilon=\frac{1}{2}\left(B\,\frac{d\phi}{dx}-W_{\phi}\right)^2+
B\,W_{\phi}\,\frac{d\phi}{dx}.
\ee
To minimize the energy, the first term in the above expression is set to zero
\be
\label{eqphi}
\frac{d\phi}{dx}=B^{-1}\,\frac{dW}{d\phi},
\ee
with minimum energy solutions given by
\be
E=B\,|\Delta W|,
\ee
where $\Delta W=W[\phi(\infty)]-W[\phi(-\infty)]$. Solutions to the
first-order equation (\ref{eqphi}) are known as Bogomol'nyi-Prasad-Sommerfeld
or BPS solutions \cite{bps,bps1}. We notice that $T^{11}$ vanishes for the BPS
solutions, as expected.

For the proton self-interaction we consider
\be
W(\phi)=\frac{1}{3}\lambda\phi^3-\lambda A^2\phi,
\ee
which gives the desirable double-well potential
\be
U(\phi)=\frac{1}{2}\lambda^2(\phi^2-A^2)^2,
\ee
where $A$ is a real and positive dimensionless parameter, $\lambda$ is also
real and has dimension length$^{-1}$. $U(\phi)$ has minima located at
${\bar\phi}_{\pm}=\pm\,A$ and the barrier height $\lambda^2\,A^4/2$.
Since the parameter $A$ gives the minima of the potential, it is directly
related to the distance between neighbor oxygens in the film network. This
allows the presence of two degenerate ground states, one with all the protons
at the position $A$, and the other with the protons at $-A$. These two states
can be represented by repetitions of the basic entities
\ben
{X\!-H\;\cdots\;Y\!-H\;\cdots}\nonumber
\\
{X\;\cdots\;H\!-Y\;\cdots\;H-}\nonumber
\een
The presence of two degenerate minima allows the appearance of solitons,
which are extended solutions with finite energy that connect the two
degenerate ground states.

To investigate the presence of soliton solutions, we notice that
in the present model the first-order equation (\ref{eqphi}) becomes
\be
\frac{d\phi}{dx}={\lambda}B^{-1}(\phi^2-A^2),
\ee
It is solved to give the kink solutions
\be
\phi(x)=A\,\tanh\left({A}{B^{-1}}\,{\bar x}\right).
\ee
Here ${\bar x}$ stands for $\lambda x$, and is dimensionless. The parameters
$A$ and $B$ determine the energy $(4/3)A^3\,B^{-1}$, and the width
$l\sim B/A$ of the soliton solutions. $A$ is related to the distance
between neighbor oxygens in the H-bond network. It is an important parameter
because under compression, conductivity and surface potential of the film
change significantly when the critical area per head group limit is
reached \cite{e1,e1a,e2,e3,e4,e4a,e5}. $B$ is an effective parameter,
that depends on the fundamental parameters $\mu_1,\mu_2,\nu_1,\nu_2$
in a specific way, imposed by the field-theoretical model here considered.
Although our model is microscopic, we can think of $B$ as a phenomenological
parameter, used to infer the width and the energy of the soliton solutions.
The soliton solutions spring in responce to nonlinear effects,
and describe proton mobility in the network. The above solutions show that
protons migrate from $A$ to $-A$, but this migration takes several
units in the chain network, given in accordance with the width $l\sim B/A$
of the solution. The main characteristics of the solution is illustrated in
Fig.~2, where  the soliton  and the asymptotic values in the chain network
is depicted. Evidently, under the presence of an external electric
field the soliton may move, giving rise to a steady current along the
direction dictated by the external electric field. The picture is similar
to that in the original work on solitons in hydrogen-bonded network
\cite{pne88}.

We consider a spatial arrangement of the molecules as in Ref.~{\cite{lco98}}. 
As the film is compressed, the area per amphiphilic unit reaches a  
critical limit, in which the distance between two consecutive COOH groups
is around $7$ {\AA}; see Fig.~1. At this point, the oxygens from the
OH and film head groups are about $2.4$ {\AA}  apart.  The hydrogen
bonds get stronger, bridging the water to the film head tightly,
which forms the network that allows the proton conduction.
Experimental data shows that the conductivity increases
with decreasing of film area. This aspect can be incorporated in the model
if we assume the energy of the kink decreases with area per amphiphilic
molecule. We do this by requiring the parameter $A$ to decrease with the
distance between the head groups. 

%%%%%%%%%%%%%%%%%%%%%%%%%% FIGURE %%%%%%%%%%%%%%%%%%%%%%%%%%%%%%%%%%%%
\begin{figure}
\centerline{\psfig{figure=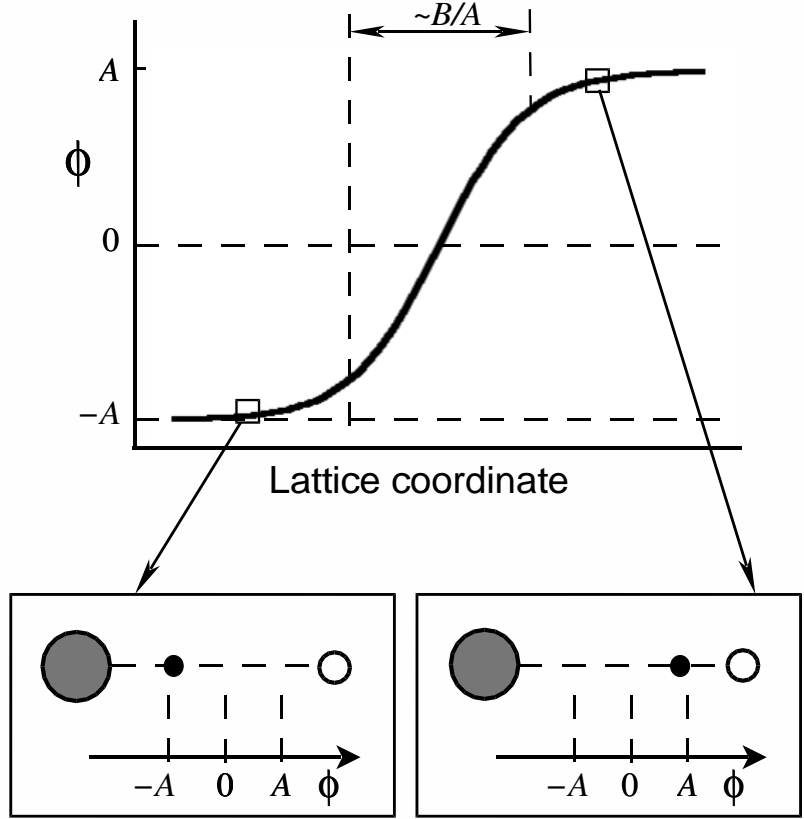,width=12.0cm}} 
\vspace{1.0cm}
\caption{Schematic view of the soliton solution, showing the corresponding
amplitude $A$ and width $l\sim B/A$. We also depict the asymptotic spots
in the chain network, to illustrate that the soliton is localized in a finite
region, involving several units of the hydrogen-bonded chain.}
\end{figure}
%%%%%%%%%%%%%%%%%%%%%%%%%%%%%%%%%%%%%%%%%%%%%%%%%%%%%%%%%%%%%%%%%%%%%

When $A$ decreases, the parameters $\nu_1$ and $\nu_2$ that control
the derivative coupling should increase, because when the head of the
amphiphilic molecule gets closer to the water molecule, the derivative
coupling should become more effective. Since $B$ is given by
$B=1-\nu_1^2/\mu_1-\nu_2^2/\mu_2$, for  increasing $\nu_1$ and $\nu_2$,
it should consequently decrease. Although we do not know the exact way
$B$ decreases, we can compare it with $A$. We can
consider, for instance, the simplest possibility in which the ratio
$A/B$ remains constant. This approximation introduces two important
consequences: i) the soliton width depends on $A/B$, so it should not
depend on the film compression; ii) the energy of the soliton depends
on $A^3/B$, so it should only depend on $A^2$. With this assumption,
our model predicts that the energy of the soliton should vary at the
rate $E/E_0=A^2/A_0^2$, for some reference value $A_0$.  As the energy
is reduced, the probability of soliton creation is increased, yielding
a higher proton conductivity in agreement with the experimental results.
In summary, the formalism presented here allows proton conductivity data
in Langmuir films to be explained.
Because the physical grounds have been established, one can now extend 
the soliton model to treat the data quantitatively, which will be the 
subject of our further investigation.
In addition, the framework may be applied to proton conductance in 
more involved  systems such as cell membranes, where proton conductance 
is believed to play an important role \cite{e1}.
Also it may be useful in the attempts to investigate effects from 
alcohols on lateral conductance \cite{yoshida}.

We thank F.A. Brito,  J.R.S. Nascimento and O.N. Oliveira Jr for discussions,
and CAPES, CNPq, FAPESP and PRONEX for financial support.

%%%%%%%%%%%%%%%%%%%%%%%%%%%%%%%%%%%%%%%%%%%%%%%%%%%%%%%%%%%%%%%%%%%%%%%%%%%%%% 

\end{document}